\documentstyle[aps,twocolumn]{revtex}
\input epsf
\input{epsf.sty}
\begin{document}
\title{Double quantum dots as a high sensitive submillimeter-wave detector}
\author{O.Astafiev \thanks{Electronic mail:
astf@mujin.c.u-tokyo.ac.jp}, S.Komiyama and T.Kutsuwa}
\address{Departament of Basic Science, University of Tokyo,
Komaba 3-8-1, Meguro-ku,Tokyo 153-8902,\\ Japan and Japan Science
and Technology Corporation, Kawaguchi City, Saitama 332-0012,
Japan}
\maketitle
\begin{abstract}
A single electron transistor (SET) consisting of parallel double
quantum dots fabricated in a GaAs/Al$_{x}$Ga$_{1-x}$As
heterostructure crystal is demonstrated to serve as an extremely
high sensitive detector of submillimeter waves (SMMW). One of the
double dots is ionized by SMMW via Kohn-mode plasma excitation,
which affects the SET conductance through the other quantum dot
yielding the photoresponse. Noise equivalent power of the detector
for wavelengths about 0.6 mm is estimated to reach the order of
$10^{-17}$ W/$\sqrt{Hz}$ at 70 mK.

\end{abstract}

 \narrowtext

\par Single-photon detection (SPD) has been achieved in the
far-infrared range by exploiting cyclotron-resonance in a
semiconductor quantum dot (QD) in high magnetic fields
\cite{Nat,PhysE,PRB}.  From the viewpoint of wide application as a
detector, it is of great importance to realize SPD without
magnetic fields as well as to expand the wavelength range. We
demonstrate here that a double QD operated as a single electron
transistor (SET) provides a novel mechanism of detecting
submillimeter waves (SMMW) with an extremely high sensitivity
close to the SPD level in the absence of magnetic fields.  The
noise equivalent power (NEP) of the detector is estimated to reach
the order of $10^{-17}$ W/$\sqrt{Hz}$ for the wavelengths of
$\lambda$ = 0.6 mm, which well exceeds reported characteristics of
conventional detectors in the relevant $\lambda$-range
\cite{SMMWdet}.

\par The mechanism is described by Figs.1 (a)-(c). The device
structure shown on Fig.1(a) is reminiscent of a lateral double-QD
SET studied earlier at different groups \cite{Kotthaus,Takahashi}.
Adjacent to the first QD (D1) that forms SET, the second QD (D2)
is placed and capacitively coupled to D1. D2 is coupled to
incident SMMW by a planar dipole antenna. As depicted in Fig.1
(b), if an electron in D2 gains an excess energy E* through the
excitation by SMMW, the excited electron escapes either to D1 or
to the electron reservoir adjacent to D2 so that the number of
electrons, $N_{2}$, in D2 decreases by one ($\Delta N_{2}$ = -1).
The electron then rapidly releases its excess energy (via phonon
emission or electron-electron interaction) relaxing to the Fermi
level, $\varepsilon_{F}$. The potential barriers, in turn, prevent
the "cold" electron from returning to D2, thereby realizing a
relatively long lifetime $\tau_{l}$ of the ionized state of D2.
Letting $C_{12}$ be an inter-QD capacitance and $C_{i}$ ($i$=1 and
2) capacitances between D$i$ and the environments, the ionization
of D2 ($\Delta N_{2}$ = -1) decreases the electrochemical
potential of D1, $\mu_{1}$, by -$\Delta\mu_{1}$ $\approx
-e^{2}C_{12}/C_{1}C_{2}$ ($e$ is the unit charge), where $C_{12}
\ll C_{1}$ and $C_{12} \ll C_{2}$ hold in the experimental
condition. This will result in the shift of SET conductance peak
by $-\Delta\mu_{1}/\varepsilon _{ch}$ $\approx -C_{12}/C_{2}$ =
$-$(3 $\sim$ 15 \%) in the sweep of $V_{G1}$, yielding a
detectable conductance change as shown in Fig.1(c), where
$\varepsilon_{ch} = e^{2} /C_{1}$ is the charging energy of D1
that determines the period of the Coulomb conductance
oscillations.

\begin{figure}
\centerline{\epsfxsize=8.0cm {\epsfbox{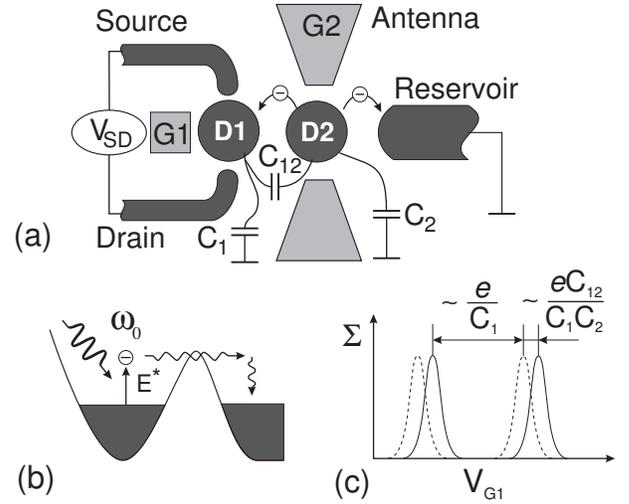}}}
\caption{Schematic representations of the SMMW photon detection.
(a) An SET consisting of parallel double QDs. (b) Ionization of
D2. (c) Conductance peak shift induced by the ionization.}
\end{figure}

\par The inset of Fig.2 schematically shows the device fabricated
through a standard lithographic technique on a
GaAs/Al$_{x}$Ga$_{1-x}$As heterostructure crystal with a mobility
of 80 m$^{2}$/Vs and a sheet carrier density of $n_{s}$ =
$2.6\times10^{15}$ m$^{-2}$ at 4.2K. Light areas indicate metal
gates deposited on top of the crystal. Negatively biasing the
gates depletes the two-dimensional electron gas (2DEG) below the
gates and forms D1, D2, the source (S), the drain (D) and the
reservoir (R). The lithographic size of each QD is 0.5$\times$0.5
$\mu$m$^{2}$, with about 200 electrons in it. The gate B12 defines
the inter-QD potential barrier. The control gate, G2, controls not
only the electrochemical potential of D2, $\mu_{2}$, but also
defines the potential barrier between D2 and R. Metal leads for G2
and B12 extend over 0.1 mm in length, forming a dipole antenna for
D2. The experiments are performed at 70mK by using a
$^{3}$He$-^{4}$He dilution refrigerator.

\begin{figure}
\centerline{\epsfxsize=7.5cm {\epsfbox{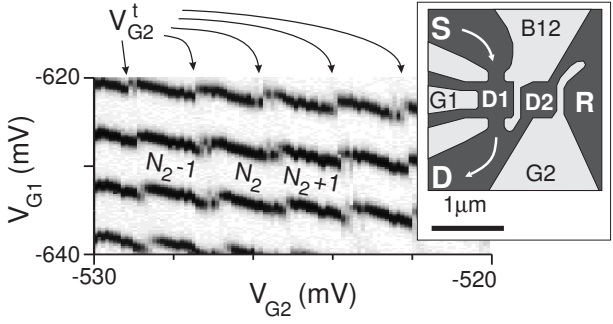}}}
\caption{Conductance peak traces on the plane of V$_{G1}$ and
V$_{G2}$. The device studied is shown in the inset.}
\end{figure}

\par The dark regions in Fig. 2 display conductance peak traces
without SMMW as a function of the bias voltages V$_{G1}$ and
V$_{G2}$ for G1 and G2, with a fixed bias voltage for B12 at
V$_{B12}$= -532 mV. As V$_{G1}$ increases, the conductance
resonance periodically occurs at the positions where $\mu_{1} =
\varepsilon_{F}$, forming a set of dark traces with a spacing
proportional to $\varepsilon_{ch} = e^{2} / C_{1}$. As V$_{G2}$
increases, each trace shows step-wise small jump in the positive
V$_{G1}$-direction at each transition point V$_{G2} ^{t}$ where
$N_{2}$ increases by one ($\mu_{2} = \varepsilon_{F}$)
\cite{Kotthaus}. This is because the change $\Delta N_{2}=+1$
causes abrupt shift of $\mu_{1}$ by +$\Delta\mu_{1}/\varepsilon
_{ch} \approx +C12/C2 \approx$ 10\% at V$_{G2}$= V$_{G2}^{t}$ .

\par As a source of SMMW, we use an n-InSb Hall device for most of
the studies and a high-mobility GaAs 2DEG Hall bar for
spectroscopic measurements, both of which emit relatively narrow
cyclotron radiation. The frequency $\omega_{c} = eB_e/m^{*}$
($m^{*}$ is the effective electron mass) is tunable by scanning
the magnetic field B$_{e}$ for the emitters \cite{PRB}. To guide
the SMMW from the emitters to the sample we use an optical scheme
similar to that previously described \cite{PRB}. The SMMW power,
$W$ , incident on the effective antenna area for D2 (about 0.2 mm
diameter) is weak, being roughly estimated to be $W \approx$ 1 fW
or $3\times10^6$ photons per second in a band width of 0.6 $\pm$
0.2 mm when the electrical input power of $P_{in}$ = 1 mW is fed
to the n-InSb emitter. The SET conductance is measured with an
ac-voltage (25 mV and 1 kHz) while SMMW are applied in a square
waveform at a lower frequency (7 Hz). We represent the
photoresponse by the difference between the dark conductance
$\Sigma_{0}$ and the conductance $\Sigma$ with SMMW, $\Delta\Sigma
\equiv \Sigma - \Sigma_{0}$, and study it via a double lock-in
technique \cite{PRB}. The wavelength of applied SMMW is in a range
0.6 $\pm$ 0.2 mm unless otherwise specified.

\par Fig. 3 (a) shows a typical photoresponse signal,
$\Delta\Sigma$, along with $\Sigma_{0}$ in a sweep of V$_{G1}$,
where V$_{B12}$=-532 mV, $W$ = 0.3 fW, and the effective
time-constant of the measurements of 1 second.  V$_{G2}$ is chosen
to be -628 mV, which is close to but higher than a nearest
V$_{G2}^{t}$. In this gate bias condition, the inter-QD coupling
($C_{12} / C_{2}$) is smaller than that for Fig.2. It follows that
the step amplitude in the conductance peak traces $\Sigma_{0}$ at
each V$_{G2}^{t}$ ($\Delta N_{2}$ = +1) is found to be smaller
than that seen in Fig.2; viz., $+\Delta\mu_{1} / \varepsilon_{ch}
\approx +C_{12} / C_{2}\approx$ 3 $\sim$ 5 \%.

\begin{figure}
\centerline{\epsfxsize=8.0cm {\epsfbox{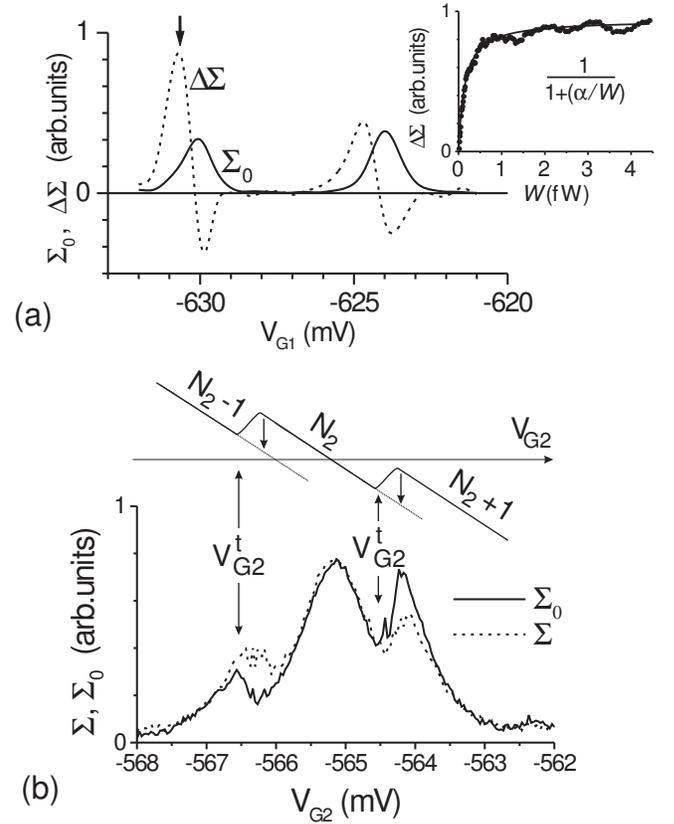}}}
\caption{(a)Conductance peaks, $\Sigma_{0}$, without SMMW (a solid
line) and the photoresponse, $\Delta\Sigma \equiv \Sigma -
\Sigma_{0}$ (a dotted line), as a function of V$_{G1}$, studied
with $W$ = 0.3 fW and V$_{G2}$ fixed at -628 mV. The inset shows
the amplitude of $\Delta\Sigma$ at the V$_{G1}$-position marked by
the arrow against the radiation intensity. (b) The dark
conductance, $\Sigma_{0}$ (a solid line), and the conductance,
$\Sigma$, under SMMW illumination (a dotted line: $W$ = 2 fW), as
a function of V$_{G2}$ with V$_{G1}$ fixed at -630 mV. A solid
line with steps in the upper space schematically represents the
conductance peak trace, while the horizontal straight line
indicates the trace along which VG2 is scanned.}
\end{figure}

\par The curve of $\Delta\Sigma$ versus V$_{G1}$ shows that the
SMMW causes the conductance peak to shift towards the negative
direction of V$_{G1}$ by 3 $\sim$ 5 \%, strongly suggesting the
ionization of D2 ($\Delta N_{2}$ = -1). Though not shown here, we
have carefully confirmed that the shape of the $\Delta\Sigma$
versus V$_{G1}$ curve is kept unchanged with increasing $W$ up to
4 fW. The negative peak shift as well as its amplitude, together
with the fact that these features are independent of $W$,
definitely indicate that $\Delta\Sigma$ arises from the switch
between the ground state ($N_{2}$) and the ionized state ($N_{2}$
- 1), as we have expected in Fig.1. On the other hand, the
amplitude of $\Delta\Sigma$, (studied at the peak position marked
by the arrow in Fig.3 (a)) linearly increases with increasing $W$
only in a limited range of weak $W$ but is saturated at higher
levels as shown in the inset of Fig.3 (a). The saturation can be
reasonably interpreted as a consequence that the rate of photon
absorption at D2 exceeds the inverse lifetime, $\tau_{l}^{-1}$, at
$W \geq$ 1 fW. The expected curve of saturation,
$\Delta\Sigma\propto (1+ (\alpha/W))^{-1}$, well accounts for the
experimental data as shown by the solid line in the inset of Fig.
3(a), where $\alpha = h\nu/\eta\tau_{l}$ with the photon energy
$h\nu$ and the quantum efficiency of photon absorption $\eta$.
Noting $h\nu$ = 2 meV and assuming $\eta = 10^{-2} \sim 10^{-3}$
[3], we roughly estimate $\tau_{l}$ = 0.1 $\sim$ 1 ms.

\par Additional evidence supporting our interpretation is
presented in Fig. 3(b), where $\Sigma$ and $\Sigma_{0}$ are shown
against V$_{G2}$ at fixed V$_{G1}$ (-630mV) and V$_{B12}$ (-530
mV). Here, $\Sigma$ is taken under constant illumination at $W$ =2
fW. The values of $\Sigma_{0}$ reach the maximum at V$_{G2}$ =
-565.2 mV, where exact conductance resonance ($\mu_{1} =
\varepsilon_{F}$) takes place. In addition, as V$_{G2}$ increases,
$\Sigma_{0}$ shows an abrupt decrease and an abrupt increase,
respectively, at the transition points V$_{G2}^{t}$ = -566.5 mV
and -564.4 mV, at which $N_{2}$ increases by one ($\Delta N_{2} =
+1$ at $\mu_{2} = \varepsilon_{F}$). When SMMW is turned on, the
conductance changes significantly on the positive sides of these
V$_{G2}^{t}$ transition points. The change is exactly the one
expected from the removal of one electron from D2 ($\Delta N_{2} =
-1$), as may be understood from the conductance peak trace
schematically illustrated on the upper part of Fig.3 (b).

\par The V$_{G2}$-range where the photoresponse occurs is limited to a
narrow interval on the positive side of V$_{G2}^{t}$. This is a
general feature found at every transition point, and is
interpreted by noting that the ionization energy of D2,
$\Delta\varepsilon = \mu_{2} - \varepsilon_{F}$, vanishes at
V$_{G2}$ = V$_{G2}^{t}$, but increases as V$_{G2}$ goes away from
V$_{G2}^{t}$ towards the positive direction. This may lead to a
rapid decrease both in $\tau_{l}$ and in the probability of
photo-ionization, restricting the V$_{G2}$-range of the
photoresponse as observed.

\par Excitation spectrum is studied by tuning the wavelength of
the cyclotron emission line ($\Delta\nu \approx$ 1.5 cm$^{-1}$)
from the GaAs/AlGaAs emitter ($m^{*}$= 0.067 $m_{0}$) over a range
5 cm$^{-1} < \nu <$ 100 cm$^{-1}$. The radiation intensity is
chosen to be in a linear response regime ($W \approx$ 0.2 fW).
Figure 4 shows the $B_{e}$-dependence of $\Delta\Sigma$ at the
peak position of the data in Fig. 3 (a), where $B_{e}$ is
converted to the frequency, $\nu = eB_{e} /2\pi m^{*}$). Distinct
resonance is found at n = 17 cm$^{-1}$ ($h\nu$ = 2 meV or $\lambda
\approx$ 0.6 mm) with a FWHM of $\Delta\nu\rm_{FWHM} \approx$ 3.5
cm$^{-1}$. We identify $\nu$ = 17 cm$^{-1}$ as the Kohn-mode
plasma resonance \cite{Heitman}, because the value agrees with the
calculation of the bare confinement potential for D2 as well as
with the extrapolation of the (plasma-shifted) cyclotron resonance
studied in our previous experiments on a similar QD (see Eq. (13)
in \cite{PRB}). We suppose that the initially excited collective
motion of electrons is very rapidly transferred to a single
electron excitation \cite{Kawabata} (within a lifetime of
(2$\pi\Delta\nu\rm_{FWHM})^{-1}$ $\approx$ 2.2 ps), so that the
excited electron with $E^{*} = h\nu$ = 2 meV escapes from D2.

\par The sensitivity of detection is extremely high as suggested
from the curve of $\Delta\Sigma$ versus $W$ in the inset of Fig. 3
(a). Furthermore, it is noted to be close to the SPD level by the
study of real-time trace of $\Sigma$. As shown in the inset of
Fig. 4, photoresponse arises as irregular conductance spikes.
Here, the data are taken with a time-constant of 3 ms at $W$ =
0.15 fW in the same gate bias condition as that for the marked
peak position in Fig. 3 (a). The density of the conductance spikes
increases with $W$. The positive spikes are ascribed to the
switches, $N_{2} \rightarrow N_{2}-1$, although individual events
of photon absorption cannot be clearly discerned because
$\tau_{l}$ (0.1 $\sim$ 1 ms) is shorter than the time-constant of
measurements. We find that the conductance spikes do not
completely vanish in the dark condition, probably, because the
sample is not perfectly shielded against the 4.2K-black-body
radiation in the present work. Noting the dark switches, we
roughly estimate NEP to be of the order of 10$^{-17}$
W/$\sqrt{Hz}$.

\begin{figure}
\centerline{\epsfxsize=8.0cm {\epsfbox{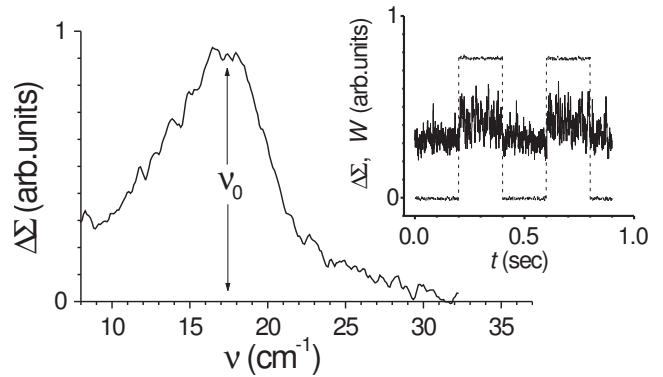}}} \caption{The
excitation spectrum of $\Delta\Sigma$, studied with $W$ = 0.15 fW
in the same gate bias condition as that of the V$_{G1}$-position
marked by the arrow in Fig. 3 (a). The inset shows a real-time
trace of the conductance at $W$ = 0.15 fW taken in the same gate
bias condition, where the square waveform indicates on-and-off of
the SMMW.}
\end{figure}

\par In conclusion, we have demonstrated ultra-high sensitive
detection of SMMW by using an SET consisting of parallel double
QDs.

\end{document}